\documentclass[oneside, a4paper, onecolumn, 11pt]{article}
\usepackage{multirow}
\usepackage[left=1.5cm,top=2cm,bottom=2cm,right=1.5cm]{geometry}
\usepackage[utf8]{inputenc}
\usepackage{graphicx} 		
\usepackage{amsmath}  		
\usepackage{eurosym}
\usepackage{longtable}
\usepackage[font=small,labelfont=bf]{caption}

\begin{document}
\begin{center}
\LARGE{\textbf{Growing structure based on viscous actuation of constrained multistable elements}
}
\end{center}
\vspace{3 pt}
\begin{center}
\large{Ezra Ben Abu$^1$,
       Yaron Veksler$^1$,
       Shai Elbaz$^1$,
       Anna Zigelman$^1$, and
       Amir D. Gat$^{1,*}$}\\
\vspace{5 pt}
\small{$^1$Faculty of Mechanical Engineering, Technion - Israel Institute of Technology, Haifa 3200003, Israel\\
$^*$amirgat@technion.ac.il}
\end{center}

\vspace{3 pt}
\begin{abstract}
Growing soft materials which follow a 3D path in space are critical to applications such as search and rescue and minimally invasive surgery. Here, we present a concept for a single-input growing multi-stable soft material, based on a constrained straw-like structure. This class of materials are capable of maneuvering and transforming their configuration by elongation while executing multiple turns. This is achieved by sequenced actuation of bi-stable frusta with predefined constraints. Internal viscous flow and variations in the stability threshold of the individual cells enable sequencing and control of the robot's movement so as to follow a desired 3D path as the structure grows. We derive a theoretical description of the shape and dynamics resulting from a particular set of constraints. To validate the model and demonstrate the suggested concept, we present experiments of maneuvering in models of residential and biological environments. In addition to performing complex 3D maneuvers, the tubular structure of these robots may also be used as a conduit to reach inaccessible regions, which is demonstrated experimentally. 
\end{abstract}


\section{Introduction}\label{sec1}
We present a growing and maneuvering multi-stable soft  material; on the sequential activation of constrained bi-stable frusta via viscous flow from a single input. The tubular structure of the robot allows it to transfer materials to unreachable regions, via an internal channel. The constraints define the final configuration of the robot, and the viscous flow allows a sequenced actuation of the straw elements, which enables the robot to follow a desired 3D path as the structure grows.

Growing materials' ability to travel along 3D paths in space is fundamental to their utility. Advancing along a 3D path and avoiding obstacles within complex environments is common in applications such as search and rescue missions~\cite{Han,gerboni2015modular} in natural or man-made disaster sites, where a survivor might be trapped below a pile of porous debris with limited oxygen or water, and minimally invasive surgery~\cite{hu2018steerable,endovascular,runciman2019soft} such as intravascular catheterization procedures. The inherently large number of degrees of freedom of these robots makes their actuation challenging. Additionally, following complex paths often requires passing through intricate narrow cavities and networks, so these robots have to be slender in order to accomplish this. Slenderness, however, typically limits a robot's ability to move forward and to turn rapidly due to the friction that develops when pushing it forward.

Growing materials, lengthening from the tip, involve no relative movement of the body with respect to the terrain. For slender growing materials with strong friction, this provides a convenient solution allowing the robot to maneuver by elongation in complex environments. One example of such robots, which are able to navigate inside a 3D maze through growing, was proposed by Hawkes et al \cite{hawkes2017soft} in the context of inverted thin-walled vessels. This concept was also extensively studied by others for various applications \cite{Siegwart,talas2020design,sadeghi2020passive,greer2020robust,sadeghi2017toward,greer2019soft,slade2017design,naclerio2018soft}.

As mentioned above, slender growing materials inherently involve an extremely large number of degrees of freedom. Thus, 
a major challenge in soft robotics is to simplify the robot's actuation, reducing the required control mechanisms. The development of strategies for efficient actuation and control of growing soft robots is essential to the advancement of the field. For instance, a single input actuation was investigated e.g., by Mosadegh et al~\cite{pneomatic_network}, who used a pneumatic-network, which consists of small channels in elastomeric materials. Yang et al~\cite{buckling} proposed the design of a ``single-unit buckling actuator'' which consists of an elastomeric structure with a ``nonbuckling center area'' connecting to several ``buckling pillars.'' Later on, Jin et al~\cite{mechanical_valve} proposed a design of multi-functional robots that operate with a single pressure input and without the need for electronic components. In particular, they utilized viscous flow and snapping arch principles, fully integrated on-board, enabling the control of the incoming airflow. Another design incorporating the interplay between bi-stability and soft actuators, was suggested by Gorissen et al~\cite{jumper} for a single-input jumping robot, where the snapping of elastomeric spherical caps upon pressurization results in a sudden release of energy which leads to a rapid jump. Flow-based sequenced actuation of multiple bi-stable elements was studied by Ben Haim et al~\cite{benHaim2019}, who provided a closed-form model for the dynamic control of multiple bi-stable hyperelastic balloons. A design for growing soft-robot with a single actuation input was proposed by Connolly et al~\cite{fiber_reinforced}, who focused on fiber-reinforced actuators, where for a given trajectory, they found the optimal design parameters for an actuator.

Here, we present a growing and maneuvering multi-stable structure based on the sequential activation of bi-stable elements via viscous flow. As the multi-stable structure, a commercially available straw composed of a sequence of conical frusta is a convenient and natural candidate. The bi-stability of the conical frusta, resulting in a multi-stability of the straw-shaped structures, was studied by Bende et al~\cite{bende2018overcurvature}, who linked the geometry and internal stress properties to multi-stable functionality in the bending and extension states. Most recently, Breitman et al~\cite{Breitman_2022}, investigated the fluid-solid-interaction dynamics occurring in a multistable straw filled with highly viscous fluid. Other multi-stable truss structures with similar properties to those of conical frusta were recently investigated by Hua and coworkers~\cite{hua2019multistable}, as well as ~\cite{wei2018design,yang2019multi}. 

The purpose of this study is to explore how viscous flow and a constrained slender multi-stable structure can be leveraged to accomplish controlled growth and maneuvering of multi-stable growing material using sequential activation of bi-stable elements. Below, we derive a theoretical description of the shape and dynamics resulting from a particular set of constraints. Models of residential and biological environments are used to experimentally demonstrate the robot's ability to perform complicated 3D maneuvers and a variety of additional operations such as heart structural intervention and fire extinguishing. 

\section{Results}

\begin{figure}[h!]
\includegraphics[width=1\linewidth]{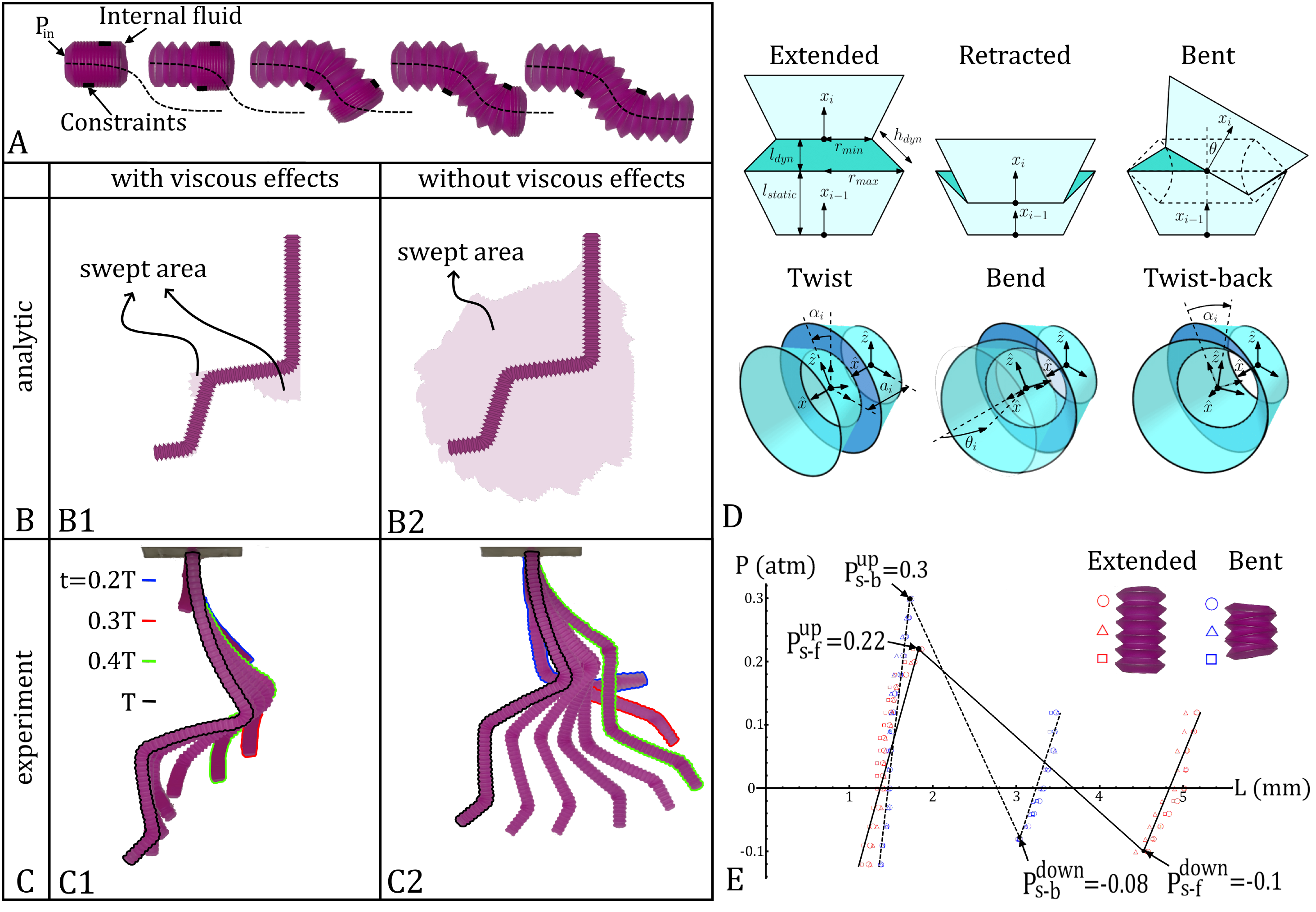}
\centering
\caption{
{\bf{Demonstration of growth of a multistable structure along with schematic sketches of all possible configurations and mechanical properties.}}
(A) The growth of a straw-like structure with constraints (marked by a black continuous lines). The sequenced opening of the bi-stable elements which results in growth along a pre-defined path (marked by a black dashed line). Panels (B1) and (B2) present kinematic simulations for sequenced growth (viscosity dominated dynamics) and unordered growth (negligible viscosity), respectively, where the pink regions denote the swept area, defined as all regions which the structure passed through during the growth process. Panels (C1) and (C2) present experimental results for actuating the configurations by a viscous internal fluid (silicone oil with viscosity of 60 Pa$\cdot$sec) at times, $t=0.2T,\ldots,T$, where $T=6$ sec, and by an inviscid internal fluid (air) at times, $t=0.2T,\ldots,T$, where $T=2$ sec, respectively. (D) Illustration of detailed descriptions for all possible element's configurations. (E) Experimental measurements of stiffness of different states and stability thresholds. Experimental data showing the internal pressure in the straw, $P$ vs. the frustum elongation, $L$, corresponding to retracted-extended snap-through (red markers) and retracted-bent snap-through (blue markers). The different markers indicate different experiments and straws. The frustum length was calculated by measuring the average elongation of 6 connected frusta.}
\label{Fig1}
\end{figure}

We suggest leveraging internal viscous flow in a single-input multistable growing material to achieve controlled sequenced actuation and growth-based locomotion in a complex 3D environment. As a model for manipulation and control of the proposed robot, we chose a straw-like structure, see Fig.~\ref{Fig1}(A), which is a common, fluid-sealed, slender multistable structure. In order to maneuver in a complex environment, it is essential to steer while elongating, which may be achieved by creating asymmetric constraints in different segments of the straw.  Such constraints can be created by using various methods, and for the current configuration polypropylene sheets were soldered in the desired regions (see methods section), marked by black lines in Fig.~\ref{Fig1}(A)), which hold one side of the straw closed. \\

\subsection{Using viscous effects to minimize the swept area via sequencing}
In Fig.~\ref{Fig1}(A), we show an example of a planar path (marked by a dashed black curve), along with the straw configuration at different times during its growth. The path determines the positions of the constraints, as well as the number of frusta that are stitched together at each location. For the current configuration, each frustum allows the steering angle of $\approx 16^{\circ}$. Using straws with a different frustum length or a different outer diameter allows various resolutions of the steering angle. We note that before the straw steers in a given direction, the front (closed frusta region) deviates from the desired path. During the propagation along the desired path, this deviation decreases, since fewer closed frusta are left in the straw. 

We define the \emph{swept area} as all locations which the soft-robot occupied during the process of growth to its final state. To follow a 3D path accurately, the swept area should be minimized, ideally to only the final 3D path. In Fig.~\ref{Fig1} panels (B1) and (B2) we show the results of planar kinematical simulations (see code in SI, as well as kinematic analysis section) of the constrained straw where the constraints' locations and the number of constrained frusta were dictated by the desired final configuration. Figs.~\ref{Fig1}  compares simulations of the swept area for both sequenced actuation (B1) and random actuation (B2). It can be seen that the swept area in Fig.~\ref{Fig1}(B1) is much smaller than in Fig.~\ref{Fig1}(B2), demonstrating the importance of a sequenced actuation to accurately grow along a predefined path. This result is also verified experimentally in Fig.~\ref{Fig1}(C1) (viscous fluid based sequenced actuation) and (C2) (air pressurization leading to random actuation). More data on the experimental parameters are available in the methods section below.\\

\begin{figure}[h!]
\includegraphics[width=1\linewidth]{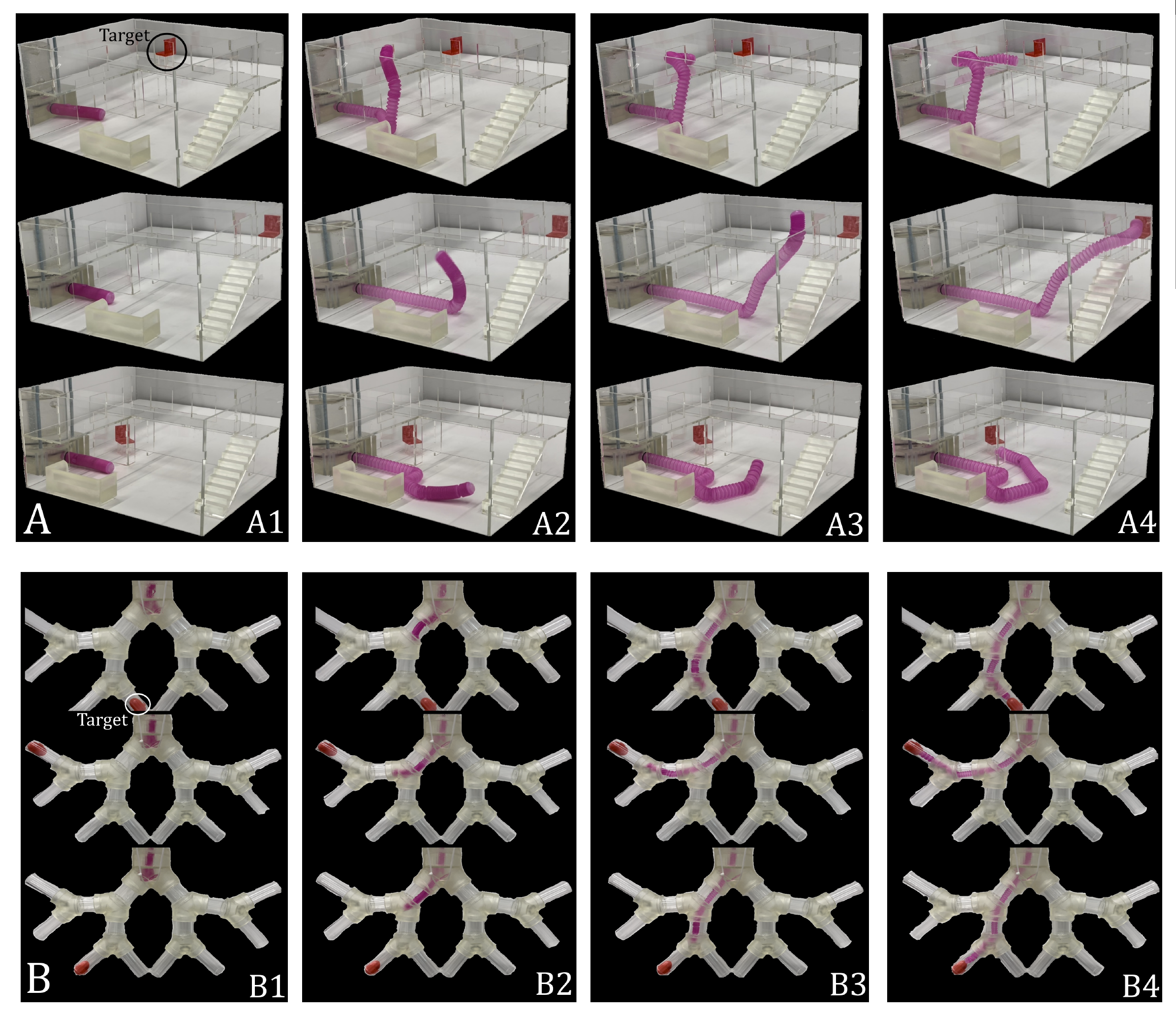}
\centering
\caption{
{\bf{Growth-based maneuvering in residential and lungs-like structures.}} Panel (A) is a residential prototype with three different geometric targets, marked by the red chair. In each series: (A1) is the initial configuration, (A2) and (A3) are intermediate stages, and (A4) is the final configuration. The final length and the duration of the growth process are in the upper row of (A) $296 \text{mm}, 8 \text{sec}$, in the middle row of (A) $304 \text{mm}, 6 \text{sec}$, and in the lower row of (A) $292 \text{mm}, 9 \text{sec}$. Panel (B) is a lungs-like prototype with three different geometric targets, marked by the red balloon. The diameter of all tubes is reduced by approximately 30\% at each junction and the opening angle of each junction is 130$^{\circ}$. In each series: (B1) is the initial configuration, (B2) and (B3) are intermediate stages, and (B4) is the final configuration. The final length and the duration of the growth process are in the upper row of (B) $313 \text{mm}, 6 \text{sec}$, in the middle row of (B) $311 \text{mm}, 7 \text{sec}$, and in the lower row of (B) $313 \text{mm}, 6 \text{sec}$.}
 \label{Fig2}
\end{figure}

\subsection{The effect of constraints on the stability threshold}
In Fig.~\ref{Fig1} panel (E) we show the pressure in the straw, $P$, vs the experimentally measured values of the frustum elongation, $L$, for experiments with and without constraints. All straws contained six frusta, and for the constraint experiments, the straw was bent in each frustum by right-left-right-left-right-left, so that overall, its elongation was forwards. We denote the upward snapping pressures by $P_{\text{s-b}}^{\text{up}}$ ($0.3$ atm) and $P_{\text{s-f}}^{\text{up}}$ ($0.22$ atm), which are the minimum pressure values needed for opening one frustum, with and without a constraint (``b'' represents constrained elongation during bending and ``f'' represents un-constrained forwards only elongation), respectively. Similarly, the downward snapping pressures denoted by $P_{\text{s-b}}^{\text{down}}$ ($-0.08$  atm) and $P_{\text{s-f}}^{\text{down}}$ ($-0.1$ atm), represent the maximum pressure values needed for closing one frustum, with and without a constraint. In all measurements, the standard deviation of the snapping pressure is below $0.01$atm.

It can be seen that, the constraints affect the threshold of stability, since $P_{\text{s-b}}^{\text{up}}>P_{\text{s-f}}^{\text{up}}$ and $P_{\text{s-b}}^{\text{down}}<P_{\text{s-f}}^{\text{down}}$. This implies that if the applied pressure is in the range $(P_{\text{s-f}}^{\text{up}},P_{\text{s-b}}^{\text{up}})$,  then only the unconstrained frusta will open and the constrained ones will remain closed and  if the applied pressure is in the range $(P_{\text{s-b}}^{\text{down}},P_{\text{s-f}}^{\text{down}})$ only the constrained frusta will close and the unconstrained ones will remain open. On the other hand if the applied pressure is equal to or greater than $P_{\text{s-b}}^{\text{up}}$, then the unconstrained frusta will open first, and afterwards the constrained ones. Furthermore, the graph in Fig.~\ref{Fig1} panel (E) presents that the elongation of the constrained frusta is reduced by a factor of 2 relative to the unconstrained ones. In addition, the ratio between the unstable regions slopes $k_{\text{s-b}}$ and $k_{\text{s-f}}$, for the constrained and unconstrained frusta, respectively, is approximately 2.5, specifically $k_{\text{s-b}}\approx-27,000\, \text{kPa}/\text{m}$ and $k_{\text{s-f}}\approx -11,000\, \text{kPa}/\text{m}$. This observation is important for modeling and controlling this system, and particularly for demonstrating that, in the case of negligible viscous effects, the frusta without  a constraint opens before the frusta with a constraint. \\

\subsection{Relating the constraints positions to the growth kinematics}
A straw-structure can be viewed as a one-dimensional lattice composed of multi-stable elements. Each element is constrained to a small set of possible states, where the states may be obtained by different geometric transformations. These transformations can generally be divided into three categories: (i) retracted element, (ii) extended element, and (iii) bent element (see Fig. \ref{Fig2}(D)). To describe the full kinematics of the entire structure, a local coordinate system is prescribed for each element. These local coordinate systems may be defined according to the Denavit-Hartenberg notation \cite{denavit1955kinematic}, where the local axial direction of the element is $x$, whereas the $z$-axis lays in the cross-section of the element (see Fig. \ref{Fig2}(D)). To transform one coordinate system into another, the following transformation matrix is used,
\begin{equation}
^{i}T_{j}=\left[\begin{array}{cc}
Rot_{3x3} & Trans_{3x1}\\
0_{1x3} & 1
\end{array}\right]\label{eq:trans_mat_def},
\end{equation}where \(Rot_{3x3}\) is a rotation matrix, \(Trans_{3x1}\) is a translation vector, and \(0_{1x3} \) is a vector of zeros.
For a serial structure, it is usually straightforward to construct the transformation matrix between two consecutive elements' coordinate systems. Hence, to transform the world coordinate system into an element's coordinate system , one must multiply all previous local transformation matrices:
\begin{equation}
^{w}T_{i}={}^{w}T_{0}\cdot{}^{0}T_{1}\cdot{}^{1}T_{2}\cdot\ldots\cdot{}^{i-1}T_{i}.\label{eq:trans_mat_series}
\end{equation}
The local transformation from element \(\left(i-1\right)\)  to element \(\left(i\right)\), denoted by \(^{i-1}T_{i}\), is comprised of a translation, \(a_{i}\), along \(x_i\) (\(Trans_{x_{i}}\left(a_{i}\right) \)) and a bending rotation with angle \(\theta_{i}\) around an axis \(u_{i}\), where the \(u_{i}-\)axis is located  in the cross-section of element $i$ and may be obtained by rotating \(z_i\) by an angle \(\alpha_{i}\) around the \(x_{i}-\)axis. To compute the local transformation matrix for straw elements, we can decompose the bending rotation into three rotations (see Fig. \ref{Fig1}(D)). First, rotating the coordinate system around \(x_i\) with angle \(\alpha_{i}\) (twist), then rotating with angle \(\theta_{i}\) around the new \(z_{i}\) (bend), and finally rotating back around the new \(x_i\) with angle \(\left(-\alpha_{i}\right)\), in order to bring the \(z_i-\)axis back to its original orientation:
\begin{equation}
^{i-1}T_{i}=Trans_{x_{i}}\left(a_{i}\right) Rot_{x_{i}}\left(\alpha_{i}\right)Rot_{z_i}(\theta_{i})Rot_{x_i}\left(-\alpha_i\right)\label{eq:local_trans_mat_decomp},
\end{equation}
where the translation and rotation matrices are defined as,
\begin{align}
Trans_{x_{i}}\left(a_{i}\right) & =\left[\begin{array}{ccc|c}
1 & 0 & 0 & a_{i}\\
0 & 1 & 0 & 0\\
0 & 0 & 1 & 0\\
\hline 0 & 0 & 0 & 1
\end{array}\right]\label{eq:translation_mat}\\
Rot_{x_{i}}\left(\alpha_{i}\right) & =\left[\begin{array}{ccc|c}
1 & 0 & 0 & 0\\
0 & \cos\left(\alpha_{i}\right) & \sin\left(\alpha_{i}\right) & 0\\
0 & -\sin\left(\alpha_{i}\right) & \cos\left(\alpha_{i}\right) & 0\\
\hline 0 & 0 & 0 & 1
\end{array}\right]\label{eq:rotation_x_mat}\\
Rot_{z_{i}}\left(\theta_{i}\right) & =\left[\begin{array}{ccc|c}
\cos\left(\theta_{i}\right) & \sin\left(\theta_{i}\right) & 0 & 0\\
-\sin\left(\theta_{i}\right) & \cos\left(\theta_{i}\right) & 0 & 0\\
0 & 0 & 1 & 0\\
\hline 0 & 0 & 0 & 1
\end{array}\right]\label{eq:rotation_z_mat}.
\end{align}
The values of the parameters \(a_{i}\), \(\theta_{i}\), and \(\alpha_{i}\) depend on the geometry of the straw element and its state. The geometrical model of a straw element assumes that each element is constructed from a static frustum of length \(l_{static}\) and a dynamic frustum of length \(l_{dyn}\), with outer radius \(r_{out}\) and inner radius \(r_{in}\) (see Fig.~\ref{Fig1}(D)). Table~\ref{table:forward_kinemetics} summarizes the values of the transformation parameters for different element states.

\begin{table}[h]
  \begin{center}
    \begin{tabular}{|l||c|c|c|} 
       \hline
       & \textbf{Retracted} & \textbf{Extended} & \textbf{Bent} \\ [0.5ex] 
       \hline
       $a_i$ \textbf{- Translation in }$\mathbf{\hat{x}}$ \textbf{direction}& \(l_{static}-l_{dyn}\) & \(l_{static}+l_{dyn}\) & \(l_{static}\) \\ 
       \hline
       $\theta_i$ \textbf{- Bending angle} & \(0\) & \(0\) & \(\Theta\) \\
       \hline
       $\alpha_i$\textbf{- Bending direction angle (twist)} & - & - & \(\in\left[0,2\pi\right)\) \\
       \hline
    \end{tabular}
    \caption{Translation and rotation values for different straw element states.}
    \label{table:forward_kinemetics}
  \end{center}
\end{table}
The value of the bending angle \(\Theta\) appearing in Table~\ref{table:forward_kinemetics} can be estimated by a 2D analysis of a straw element. Assuming the outer and inner radii of the element are constant, the bending angle can be found by using the cosine rule (see Fig.~\ref{Fig1}(D)):
\begin{equation}
\Theta=\arccos\left(\frac{r_{out}^{2}+r_{in}^{2}-h_{dyn}^{2}}{2\cdot r_{in}\cdot r_{out}}\right)\label{eq:bending_angle},
\end{equation}where \(h_{dyn}\) is the side-length of the dynamic frustum. Assuming \(h_{dyn}\) remains constant for all element states, its value can be calculated for an element in the extended state (see Fig.~\ref{Fig1}(D)):
\begin{equation}
h_{dyn}^{2}=\left(r_{out}-r_{in}\right)^{2}+l_{dyn}^{2}\label{eq:dyn_side_len}.
\end{equation}

A kinematic simulation of the entire straw structure was created by combining the forward kinematic computations for all straw elements, as described in equations~\eqref{eq:trans_mat_def}--\eqref{eq:rotation_z_mat}, using the parameter values for different element states as given in Table~\ref{table:forward_kinemetics} and equations~\eqref{eq:bending_angle}--\eqref{eq:dyn_side_len}. This simulation was then utilized to create the swept area domains for the two scenarios, which were shown in Fig.~\ref{Fig1}(B1) and Fig.~\ref{Fig1}(B2). The swept area in Fig.~\ref{Fig1}(B1) illustrates a sequenced elements' elongation, where all elements were transformed from a retracted state to their final state. The swept area shown in Fig.~\ref{Fig1}(B2) was obtained by taking a union of 200 domains found from random-order element activation.

\begin{figure}[ht]
\includegraphics[width=1\linewidth]{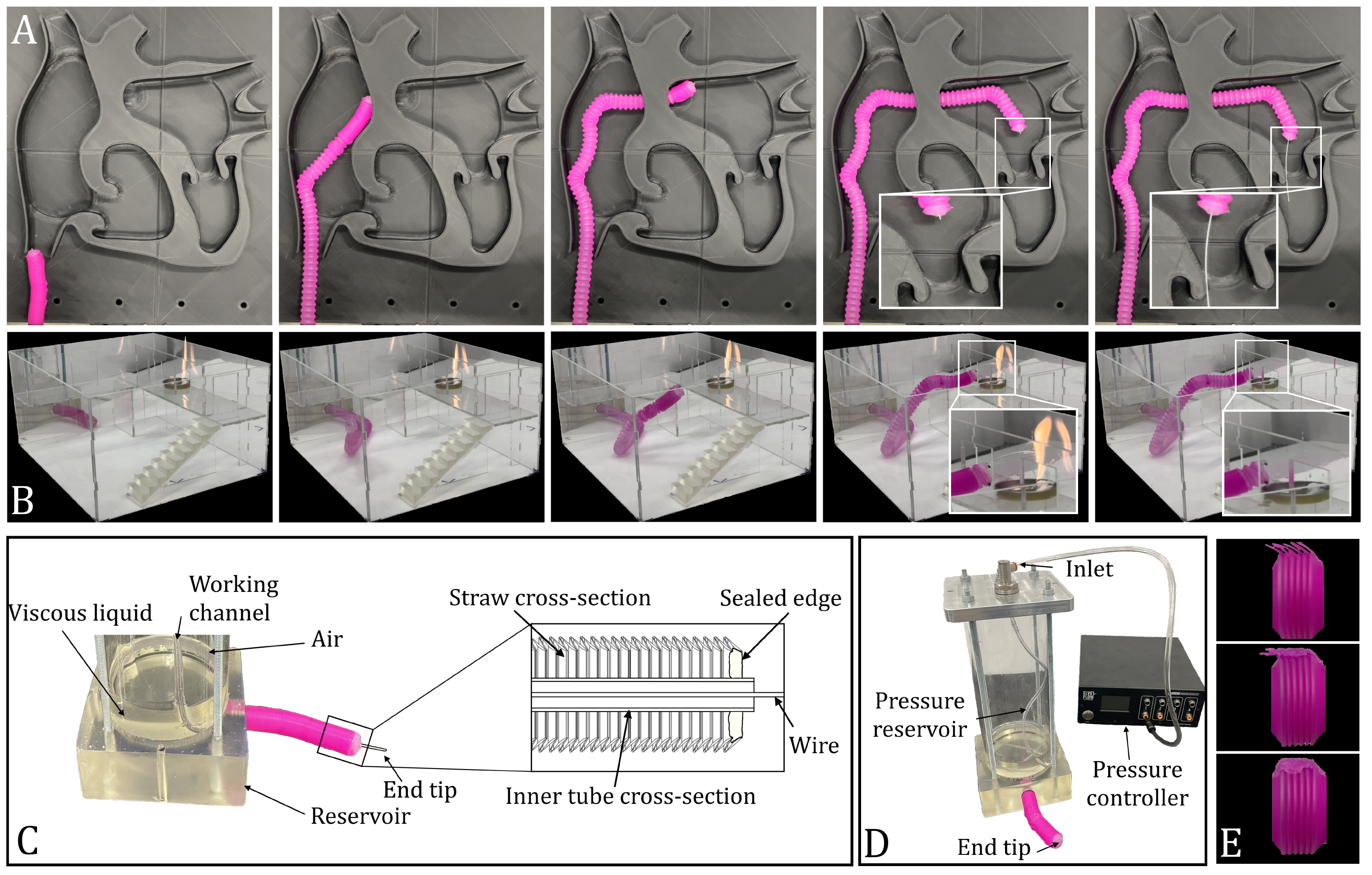}
\centering
\caption{
{\bf{Maneuvering in complex environments and transferring materials to unreachable regions, via an internal channel.}} (A) Maneuvering in a heart-like structure simulating structural heart intervention. (B) Maneuvering in a residential setting and transferring water to a specific location.  (C) A part of the experimental setup with an enlarged view of the ``end tip,'' where its cross-sectional illustration is shown in the inset. (D)  A view of the general experimental setup which consists of a pressure controller connected by two channels to the growing material. The first channel controlled the pressure in a fluid reservoir and the second channel is used as the working channel. (E) The constraints fabrication process.} \label{Fig3}
\end{figure}

\subsection{Demonstration of maneuvers and operations in complicated 3D environments}
In Figs.~\ref{Fig2}(A) and~\ref{Fig2}(B), we demonstrate two possible applications of the suggested concept of a controlled single-input growing material with viscous internal fluid, which maneuvers inside a 3D complex environment (a residential model and a model of human lungs-like structure). In both cases, we performed several experiments with different goal positions and recorded the straw growth. In the experiments presented below, the number of constrained frusta is in the range $6-16$, and the average elongation was approximately 440\% of the resting length. In these demonstrations, the growing multi-stable robot was able to follow desired paths which included passing through thin tubes and corridors, junctions, steep corners, narrow gates, dodging obstacles, overcoming gravity, and changing the plane of motion.  In Fig.~\ref{Fig2}(A), we show locomotion via growth in a residential model for three different final goal locations (marked by a red chair). It can be seen that the number of bends slightly affects the final length, e.g., when the number of bends is 16 (with overall steering angle of 256$^{\circ}$), the length decreases from 325 mm (which is the maximal length of the straw) to 296 mm (meaning that in this case the overall steering angle of 256$^{\circ}$ results in the elongation decrease of approximately 10\%). In Fig.~\ref{Fig2}(B), we demonstrate locomotion by growth in a lung-like model for three final goal locations (marked here by a red balloon).  At each junction one tube splits into two different tubes, thus at generation $n$, ($n=1,2,3,\ldots$), there are $2^n$ tubes, which increases the complexity of the maneuvering till reaching the goal as the multi-stable structure grows and gets closer to the destination. The diameter in each generation is reduced by approximately 30\%, which complicates and restricts the maneuvering abilities. In our case, $n=3$, and at the last generation the inner tubes' diameter is 150\% of the maneuvering robot's diameter.

In Fig.~\ref{Fig3} we show the ability of the proposed soft-robot to maneuver and then to perform various operations, such as heart structural intervention in the heart-like structure (see Fig.~\ref{Fig3}(A)) and putting out the fire in a residential model (see Fig.~\ref{Fig3}(B)). This figure visualizes that beyond maneuvering in a 3D complex environment and reaching the goal, the soft-robot is capable of transferring various materials such as water or medical equipment. For a detailed experimental setup including the robot's cross-sectional illustration see Fig.~\ref{Fig3}(C), where note that the working channel is an inner tube attached to the sealed end tip of the straw allowing to transfer the above mentioned materials. The inner tube is sealed and isolated from the internal pressure inside the reservoir, and is flexible enough to move with the material as it grows.

\section{Concluding remarks}
Natural phenomena as well as many engineering applications involve geometries that are narrow and complex. Maneuvering in such confined and intricate 3D environments is exceptionally challenging for conventional rigid robots. In this paper, we presented a new concept for a single-input growing material that can steer and elongate along a predefined path, as well as perform a variety of operations. In the proposed class of growing materials, we apply physical phenomena (specifically the interaction between viscosity and multi-stability) to achieve controllable locomotion, enabling single-input control. The proposed growing material is a multi-stable structure with constraints located at positions computed from the kinematic analysis of the desired 3D path. We demonstrated the feasibility of the suggested concept for various scenarios. In all cases, the growing multi-stable structure was able to follow the desired path including through narrow tubes and corridors, junctions, steep corners, narrow gates, dodging obstacles, overcoming gravity, and changing its plane of motion.

Apart from viscous-based sequencing, the growing material can be controlled via utilizing the different threshold pressures for the opening of straight and bent frusta (this concept is related to ideas explored by Peretz et al. \cite{peretz2020underactuated} and Melancon et al \cite{melancon2022inflatable}). For viscous-based sequencing, the fluidic pressure propagation determines the rate and order of opening of the frustum. The frusta are opened or closed, depending on the sign of pressure, from the inlet toward the closed end of the straw without distinction between frusta with and without constraint. In contrast, in the case of negligible viscous effects under positive pressure, we observe that frusta without a constraint open first and those with a constraint close first when activated with negative pressure. This enables additional modes and geometries of the single-input growing robot.

\section{Materials and methods}

\subsection{Research objective and design}
This study aims to utilize viscous fluid interacting with multistable elastic structures to construct a simple growing material, which is capable of performing maneuvers in complex 3D environments with obstacles and junctions, such as human lungs or natural disaster sites.  As a prototype for the elastic multi-stable structure, we used a commercially available straw to which we attached constraints at specific locations. Fabrication of the constraints (see ~\ref{Fig3}(E)) is accomplished with a standard soldering machine. In order to determine the constraint locations, we use kinematic analysis based on the desired dynamics and final configuration of the robot.

\subsection{Fabrication of experimental setup}
We used a straw made from Polypropylene, which is a relatively inexpensive and convenient material for production~\cite{ROY2021128234}. In addition it has a relatively high Young's modulus ($E=1.3\, \text{G}\cdot \text{Pa}$) which allows to fabricate a sufficiently reliable structure. In our experiments we employed a straw with 78 frusta. At atmospheric conditions, the maximal and minimal lengths of the straw are 415 mm and 90 mm, respectively. Furthermore, the outer and the inner diameters of the straw are 19 mm and 13 mm, respectively. Generally, straw-like structures with a wide range of outer diameters (6-200 mm) are commercially available. Moreover, even smaller straws, with an outer diameter of 2 mm, can be fabricated by using standard methods. 

The straw was directly attached to a fluid reservoir using a thread which is a part of the straw end. To prevent penetration, the other thread located at the second end of the straw was removed and sealed with a hot glue. In experiments with viscous fluid, the straw was filled with silicon oil (viscosity of 60 Pa$\cdot$sec). The direct attachment between the straw and the fluid reservoir allows to minimize the pressure losses, and therefore increases the efficiency of the system. The fluid reservoir was connected to a pressure controller (ELVEFLOW OB1 MK3+), which in turn was connected to a compressor (CompAir L07). In all experiments, the pressure controller was adjusted to 2.5$\pm$0.01 atm and the temperature was kept at room temperature. 
In Fig.~\ref{Fig1} the straw elongation was captured by video, in 4K resolution and 240 fps, whereas in Fig.~\ref{Fig2}(A) and (B), the straw elongation was captured by video, in 1080p resolution and 30 fps, until it reached the goal.

First, after mapping the path, by the aid of our theoretical model the locations of the constraints can be found. Then the constraints are marked on the straw, on the same sides of the straw as the desired turns, for example when a right turn is to be performed, the constraint  should be inserted on the right side of the straw. To elongate straight forwards no constraints are needed. Then, the constraints were inserted in the appropriate zones by spreading Polypropylene sheets between the neighboring frusta and soldering them together. In order to guarantee that the constraints will be perfectly adhered, the soldering machine temperature was set to 200$^{\circ}$C, which is 25\% higher than the melting point of Polypropylene. Inserting a constraint on a single frustum results in a steering angle of approximately 16$^{\circ}$, thus knowing in advance the desired steering angle of the maneuver, the number of adhered frusta is determined. 

The lungs-like structure (see Fig.~\ref{Fig2}(B)) consists from Perspex tubes of length 100 mm and with inner diameters of 54, 40, and 30 mm and special connectors (printed using SLA 3D printer -Form 3B+) that fit the tubes' diameters. The inner diameter of the inlet connector is 80 mm. The diameter of all tubes is reduced by approximately 30\% at each junction and the opening angle of each junction is 130$^{\circ}$. The lungs-like structure and the straw are attached to the bottom of the fluid reservoir. The heart-like structure (see Fig.~\ref{Fig3}(A)) was fabricated by combining four components which were printed with FDM printer (RAISE3D pro2).

The residential model (model of a small house, see Figs.~\ref{Fig2}(A) and~\ref{Fig3}(B)), was created from acrylic sheets by using a laser cutting machine (Makeblock Laserbox 40W), where the furniture was printed using SLA 3D printer (Form 3B+). The outer dimensions of the house are 250$\times$300$\times$150 mm$^3$. The height of each floor is 75 mm and the gates' dimensions are 40$\times$40 mm$^2$ and 40$\times$70 mm$^2$. The fire setup was fabricated by using a small metal can which was filled with paper soaked with IPA, and then the fire was ignited. 

The pressure reservoir (see Fig.~\ref{Fig3}(D)) used in all of the maneuver experiments was made from perspex tube, aluminium cover on top of it, and SLA printed cover on its bottom. For experiments shown in Fig.~\ref{Fig2}(A) and (B), we used the upper cover with only one pressure inlet, whereas for the experiments shown in Fig.~\ref{Fig3}(A) we added an additional input inlet (working channel) for performing various operations such as inserting a wire for heart structural intervention or using the channel as a hydrant for splashing the water on fire. This working channel was sealed and isolated from the internal pressure inside the reservoir.

{\textbf{Author contributions:}}
 A.D.G. and E.B.A. conceived the research subject. E.B.A constructed the experimental setup and conducted the experiments. Y.V. performed the theoretical analysis and numerical computations. E.B.A. analyzed the experimental data. E.B.A., Y.V., S.E., A.Z., and  A.D.G. wrote the paper.


 \subsection*{Acknowledgements}\label{sec4.11}
 This work was supported by the Ministry of Energy of Israel.








\end{document}